\documentclass[prl,amssymb,amsmath,twocolumn,floatfix,letterpaper]{revtex4}
\usepackage{hyperref}
\usepackage{graphicx}
\usepackage{psfrag}
\begin{document}

\title{Two-parameter scaling law of the {A}nderson transition}

\author{ Viktor Z. Cerovski }
\affiliation{ Institut f\"ur Physik, Technische Universit\"at, D-09107
Chemnitz, Germany. } 

\begin{abstract}
It is shown that the Anderson transition (AT) in $3d$ obeys a two-parameter
scaling law, derived from a pair of anisotropic scaling transformations, and
corresponding critical exponents and scaling function calculated, using a
high-precision numerical finite-size scaling study of the smallest Lyapunov
exponent of quasi-$1d$ systems of rectangular cross-section of $L\times\delta
L$ atoms in the limit of infinite $L$ and $\delta L < L$, for $x=\delta L/L$
ranging from $1/30$ to $1/4$. The second parameter is $x$, and there are two
singularities: apart from the two-parameter scaling describing AT for $x>0$,
corrections to scaling due to the irrelevant scaling field diverge when $x\to
0$, and the corresponding crossover length scale is also estimated.
Furthermore, results suggest that the signatures of the AT in $3d$ should be
present also in $2d$ strongly localized regime.
\end{abstract}

\date\today

\maketitle

One of the long-standing open fundamental problems of the physics of
quantum-mechanical disordered systems is the quantitative description of the
metal-insulator transition induced by the Anderson localization of electronic
eigenstates~\cite{Anderson58}, known as the Anderson or
localization-delocalization transition (AT).  The subsequently developed
scaling theory of localization of Abrahams et al.~\cite{Abrahams79} (STL)
proposed that: (i) in $3d$ the transition is a continuous phase transition with
only one relevant scaling variable (which became known as the single- or
one-parameter scaling hypothesis), (ii) the lower critical dimension is $2$
where all states are localized for arbitrarily small finite disorder strength,
and (iii) the transition can be described in terms of the scaling law of the
disorder-averaged dimensionless conductance $g$ that depends only on $g$, i.e
$d\ln g/d\ln L = \beta_d(g)$.  The theory is based on the renormalization-group
(RG) considerations of the Thouless expression for $g$, and an additional
calculation showing that $\beta_{d=2}(g)<0$, which was corroborated by a more
detailed self-consistent study~\cite{Vollhardt80} based on a resummation of the
perturbation theory for weak disorder~\cite{Langer66}.

The discovery of the universal conductance
fluctuations~\cite{Lee85,Mesoscopic91} showed that $g$ is not a self-averaging
quantity, and therefore scaling of its whole distribution has to be studied.
The STL nevertheless survived in the sense that one can still find a single
parameter $g$ that characterizes scaling properties of the whole distribution,
and there is a compelling evidence that the scaling properties of the
distribution itself in the critical region of the transition are still governed
by a single-parameter~\cite{Shapiro86}.

Numerical studies of the transition using the transfer-matrix method (TMM)
\cite{Pichard81} suggested a possibility that $\beta_{d=2}(g)=0$ for a finite
amount of disorder, which would correspond to the existence of a line of
critical points for disorder weaker than a certain finite disorder strength,
but subsequent studies showed that the dependence of localization length on the
disorder strength in $2d$ is in a quantitative agreement with analytical
results~\cite{MacKinnon81}, 

High precision numerical calculations of the critical exponent $\nu$ describing
the divergence of the localization length $\xi$ at the critical point in $3d$,
however,  give $\nu\approx 1.54$ \cite{MacKinnon94,Slevin99}, in sharp contrast
with $\nu=1$ obtained from the self-consistent analytical
calculations~\cite{Abrahams79,Vollhardt80}.

Theoretical breakthrough was made by Efetov~\cite{Efetov83}, who introduced
supersymmetry to calculate disorder-averaged products of Green's functions, and
provided a theoretical framework that, among other results, allows calculation
of $\nu$ beyond the self-consistent approach, although technical difficulties
with the $\epsilon$-expansion do not yet permit accurate estimate of $\nu$
despite the considerable theoretical progress~\cite{Hikami81}, and currently
$\nu$ is most accurately determined using TMM.

This Letter present three main results:
(1) there is an additional scaling parameter $x$ in $d=3$ describing the
thickness to width ratio of long quasi-$1d$ wires, and the corresponding
two-parameter scaling law and critical exponents are estimated numerically; (2)
results are in agreement with (ii) and additionally show that there should be
possible to see signatures of $3d$ transition in $2d$;  (3) the description of
the transition in terms of the $\beta$-function depending {\it only} on $L$ is
incomplete due to the geometric nature of both $x$ and $L$.

The starting point is the Anderson model~\cite{Anderson58}:
\begin{equation}
  H = \sum_i\epsilon_i|i\rangle\langle i| + t
      \sum_{\langle ij\rangle}(|i\rangle\langle j| + |j\rangle\langle i|),
\end{equation}
where $\epsilon_i$ represents the impurity energy at site $i$. $\epsilon_i$
is randomly, independently and uniformly distributed in $[-W/2,W/2)$; $t$ is
the hopping integral of electron (set to 1), $\langle i j\rangle$ denotes
that the hopping takes place only between the nearest-neighbors of the simple
cubic lattice, and the Fermi energy is set to 0 (band center).

The geometry studied is that of the quasi-$1d$ slabs of $L\times\delta L\times
M$ atoms with $M\gg L$, ratio $x=\delta L/L$, open boundary conditions (b.c) in
$\delta L$ direction and periodic b.c in $L$ direction.  Similar geometry of
cubic samples of $L_0\times L_0\times (q L_0)$ atoms was studied in
Ref.~\cite{Potempa98}, where authors found that the critical disorder strength
$W_c$ is approx.~independent of the shape and that $g$ becomes strongly
suppressed for $q\ll 1$ and $q\gg 1$ which is consistent with, respectively,
the quasi-$2d$ and quasi-$1d$ character of samples in these cases.

The scaling properties of AT in $3d$ are studied by the standard calculation of
the smallest Lyapunov exponent $\gamma$ of transfer matrices of long quasi-$1d$
samples~\cite{MacKinnon94,Slevin99}.
The inverse of $\gamma$ is the largest length scale in the problem, which is
identified with the correlation length $\xi=1/\gamma$.  The usually studied
quantity is the rescaled correlation length $\Lambda$ defined as
$
 \Lambda(L,W)\equiv (L \gamma(L,W))^{-1}
$

The finite-size scaling analysis of $\Lambda$ gives the scaling properties of
$3d$ systems in the thermodynamic $L\to\infty$ limit, by considering how
$\Lambda(L,W)$ changes under the RG transformation $\mathcal{R} : L\mapsto b'
L, \delta L\mapsto b' L$ \cite{Cardy96}.  The corresponding scaling law,
including the corrections to scaling due to one irrelevant field was considered
in the context of AT first by Slevin and Othsuki~\cite{Slevin99}, who were able
to numerically show that 
\begin{equation}\label{eq:one}
 \Lambda(L,w)=F(L^{1\over{\nu}}\psi(w),L^{y}\phi(w)),\; w=\frac{W-W_c}{W_c},
\end{equation}
where 
$\nu>0$ and $y<0$ are critical exponents associated with, respectively, the
relevant and irrelevant scaling fields $\psi$ and $\phi$, and all functions are
analytic.  This is achieved by fitting numerically obtained values of $\Lambda$
with the truncated expansions of $F,\phi,$ and $\psi$, while the error-bars are
estimated via calculation of $95\%$ confidence intervals that can be done
either by the bootstrapping method~\cite{Slevin99} or by a direct calculation of
projections of the confidence region~\cite{Cerovski07b}.

In the case when there is an additional parameter $x$, we can repeated the
above procedure for several values of $x$, 
\begin{eqnarray}\label{eq:x}
 \Lambda(L,w,x)=F(L^{1\over{\nu}}\psi(w,x),L^{y}\phi(w,x),x),
\end{eqnarray}
where $\nu, y, W_c$ may in general also depend on $x$.  The principal difference
between Eq.~(\ref{eq:one}) and Eq.~(\ref{eq:x}) is that $F,\psi$ and $\phi$ do
not have to be necessarily analytic in $x$.

Expansion in the second argument of $F$ gives:
\begin{eqnarray}\label{eq:x-expanded}
 \bar\Lambda(L,w,x)&=&F_\pm(L^{1\over\nu}\psi(w,x),x),\;
 \bar\Lambda=\Lambda-\Delta F,\\
 \Delta F &=&\sum_{n=1}^\infty F_n(L^{1\over\nu}\psi(w,x),x)\phi(w,x)^n L^{n y},
\end{eqnarray}
where $F_\pm$ represents the universal part describing AT, and $\Delta F$ are
corrections to scaling due to $\phi$.  These vanish for large $L$ because $y<0$
but are important for a quantitative description of AT, including precise
determination of all of the relevant parameters~\cite{Slevin99,Cerovski07b}.

\begin{table*}
\begin{ruledtabular}
\begin{tabular}{c c c c c   c c   c c c}
$x$&$\delta L$&$W$&$N_d$&$N_p$&$\chi^2$&$Q$ & $W_c$ & $\nu$ & $y$\\
\hline
$1  $ &$[4,14]$&$   [15,18]$&  $427$ &$10$& $439.1$& $0.2$ &$16.46(39,55)$&$1.60(55,65)$&$-1.39(1.16,1.62)$\\
$3/4$ &$[3,15]$&$   [15,18]$&  $304$ &$10$& $282.4$& $0.7$ &$16.52(46,60)$&$1.62(57,66)$&$-1.13(0.98,1.29)$\\
$1/3$ &$[1,4]$&$   [15,18]$&   $427$ &$10$& $418.2$& $0.5$ &$16.52(42,63)$&$1.54(48,61)$&$-1.00(0.90,1.10)$\\
$1/4$ &$[1,7]$&$   [15,18]$&   $427$ &$10$& $436.9$& $0.2$ &$16.48(42,54)$&$1.55(51,60)$&$-1.00(0.97,1.04)$\\
$1/7$ &$[1,5]$&$[14.25,18]$&   $380$ &$11$& $370.0$& $0.5$ &$16.61(47,77)$&$1.52(39,67)$&$-0.89(0.84,0.95)$\\
$1/10$&$[1,6]$&$[14.5,18]$&    $370$ &$11$& $365.7$& $0.4$ &$16.62(50,75)$&$1.52(41,66)$&$-0.88(0.84,0.93)$\\
$1/15$&$[1,4]$&$[14.5,18.25]$& $304$ &$11$& $298.2$& $0.4$ &$16.58(23,95)$&$1.62(41,87)$&$-0.90(0.80,0.99)$\\
$1/20$&$[1,4]$&$[14.5,18.25]$& $304$ &$11$& $287.3$& $0.6$ &$16.59(27,99)$&$1.57(37,89)$&$-0.90(0.80,1.00)$\\
$1/30$&$[1,4]$&$[14.5,18.5]$ & $324$ &$11$& $350.3$& $0.1$&$16.50(27,80)$&$1.59(51,69)$&$-0.93(0.85,1.01)$\\
\end{tabular}
\end{ruledtabular}
\caption{Input parameters of the simulation: $x$, thickness $\delta L$, the
number of points $N_d$, the number of parameters $N_p$; Output values: the
minimal least-square deviation $\chi^2$ obtained, the quality of fit parameter
$Q$; and the values of the fitting parameters: the critical disorder strength
$W_c$, the critical exponent $\nu$ and the irrelevant exponent $y$.  All
$\Lambda$'s are calculated with $0.1\%$ relative accuracy except for the
largest systems studied ($x=1/30, \delta L=4$), which is calculated with
accuracy of $0.12\%$.  Error-bars are $95\%$ confidence intervals.
}\label{table}
\end{table*}

Figure 1 shows the typical behavior of $\Lambda(L,W)$ and the corresponding fit
for small constant $x$ and several $\delta L$ starting with 1.  As $L=\delta
L/x$ increases, $\Lambda(L,W)$ at first decreases uniformly in $W$ (since for
small $\delta L$ system is close to being $2d$) but with further increase of
$L$ a characteristic behavior for the continuous transition begins to develop,
since for large $\delta L$ and constant $x$ system is $3d$.

\begin{figure}
\begin{center}
\includegraphics*[width=3.25in,draft=false]{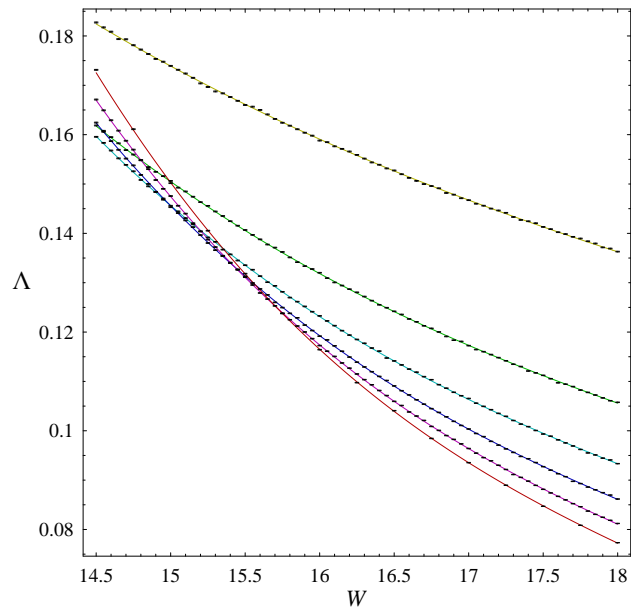}
\end{center}
\caption{
Values of $\Lambda(L,W)$ for $x=1/10$. The corresponding fits are represented
with lines for $\delta L =1,...,6$ (going from the slowest changing curve to
the fastest).}
\label{fig:LAMBDA}
\end{figure}

Table~\ref{table} summarizes the parameters and the results of the numerical
simulation using the methods described above for several values of $x$.
Results presented in the Table suggest that $\nu$ is approx.~independent of $x$
and in agreement with Ref.~\cite{MacKinnon94,Slevin99,Cerovski07b}, which
supports its universality.  Similar can be said for $W_c$, in agreement with
Ref.~\cite{Potempa98}, while $y$ becomes approx.~constant for $x\lessapprox
1/4$.

To better understand the scaling properties of AT w.r.t $x$, I consider an
additional scaling transformation, $\mathcal{R}_x:x\mapsto b x$ under constant
$L$ (this is equivalent to scaling only the thickness $\delta L$), and introduce
a scaling field $\tilde\psi$ that depends on $x$ but not on $w$:
\begin{equation}\label{eq:onex}
 \bar\Lambda = F_\pm(L^{1\over\nu}\psi(w,x),\tilde\psi(x)),
\end{equation}
where $W_c, \nu$ and $y$ are assumed to be constant.
The scaling law of $F_\pm$ w.r.t $x$ can be derived assuming that functions
scale under $\mathcal{R}_x$ in the general way~\cite{Cardy96}:
\begin{equation}\label{eq:Rxa}
 \mathcal{R}_x : F_\pm\mapsto b^{y_1}F_\pm,\;\psi\mapsto b^{y_2}\psi,\;
                 \tilde\psi\mapsto b^{y_3}\tilde\psi,
\end{equation}
Iterating $\mathcal{R}_x$ a finite number of times in the standard manner gives
the two-parameter scaling law:
\begin{equation}\label{eq:two-universal}
 \bar\Lambda = x^\alpha\mathcal{F}_\pm(L^{1\over\nu}x^\mu\psi(w)),
\end{equation}
where $\alpha\equiv -y_1/y_3, \mu\equiv -y_2/y_3$.

At first it seems that Eq.~(\ref{eq:two-universal}) cannot be correct in the
case of AT for small $x$ because numerical results give that the transition
takes place at approx.~const.~$W_c$, and if $W_c$ would remain constant for
smaller $x$ as well, Eq.~(\ref{eq:two-universal}) would imply that transition
persists when $x\to 0$ (regardless of the value of $\alpha$), where one must
find strongly localized states instead.  

\begin{figure}
\begin{center}
\includegraphics*[width=1.65in]{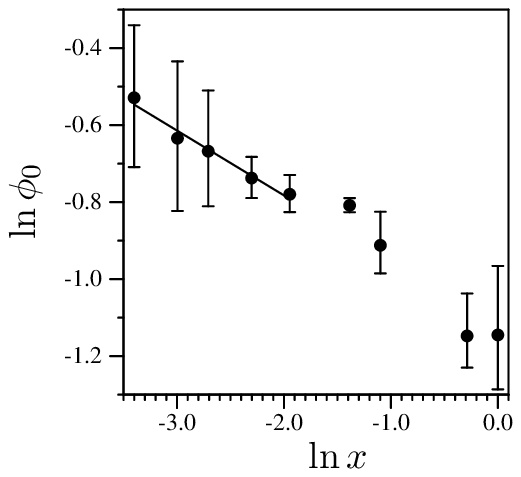}
\includegraphics*[width=1.65in]{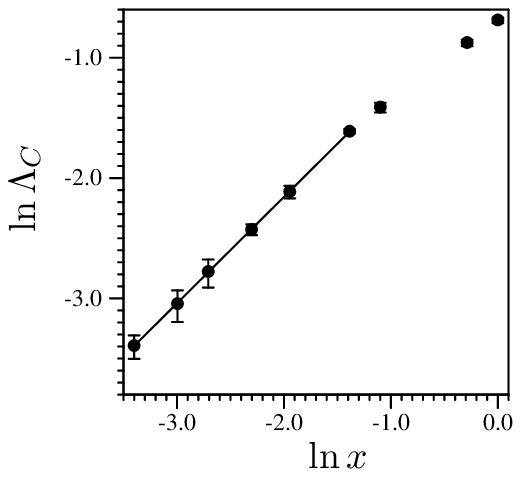}
\end{center}
\caption{Dependence of $\phi_0(x)$ (left panel) and $\Lambda_c(x)$ (right
panel) on $x$. The solid line in both panels is the least-square linear fit. 
Error-bars are $95\%$ confidence intervals.}
\label{fig:phi0}
\end{figure}

This is resolved if we notice that $\Delta F$ is also divergent as $x\to 0$,
and that there is a characteristic crossover length-scale $L_c(x)$ such that
for $L\gtrsim L_c(x)$ corrections to scaling become small, but $L_c(x)\to
\infty$ when $x\to 0$. The source of the divergence for $W\approx W_c$ is in
$\phi(x,w=0)=\phi_0(x)$, and the left panel of Fig.~\ref{fig:phi0} shows that
$\phi(x)\propto x^\delta$ for $x\lessapprox 1/7$, with $\delta = -0.17(05,29)$.
$L_c(x)$ is determined from the condition $L_c(x)^y \phi(x) \sim 1$, giving
\begin{equation}\label{eq:Lc}
  L_c(x)\sim x^{-{\delta\over y}}.
\end{equation}
A more detailed discussion of the exponent $\delta$ will be carried out
elsewhere since our main interest here is in the two-parameter scaling law
Eq.~(\ref{eq:two-universal}) of AT.

\begin{figure}
\begin{center}
\includegraphics*[width=3.25in,draft=false]{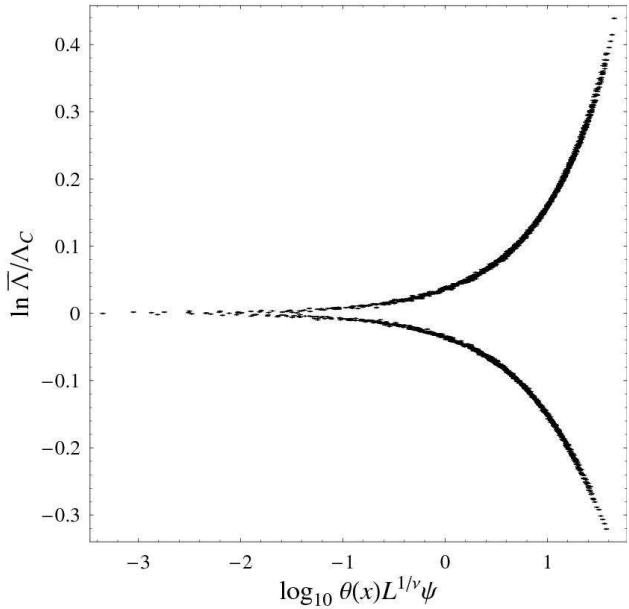}
\end{center}
\caption{Collapse of $\bar\Lambda(L,W,x)$ onto the universal function
$\mathcal{F}_\pm$, for 2109 points from Table~\ref{table} corresponding to
$x\le 1/4$. 
}
\label{fig:F}
\end{figure}

\begin{figure}
\begin{center}
\includegraphics*[width=1.52in]{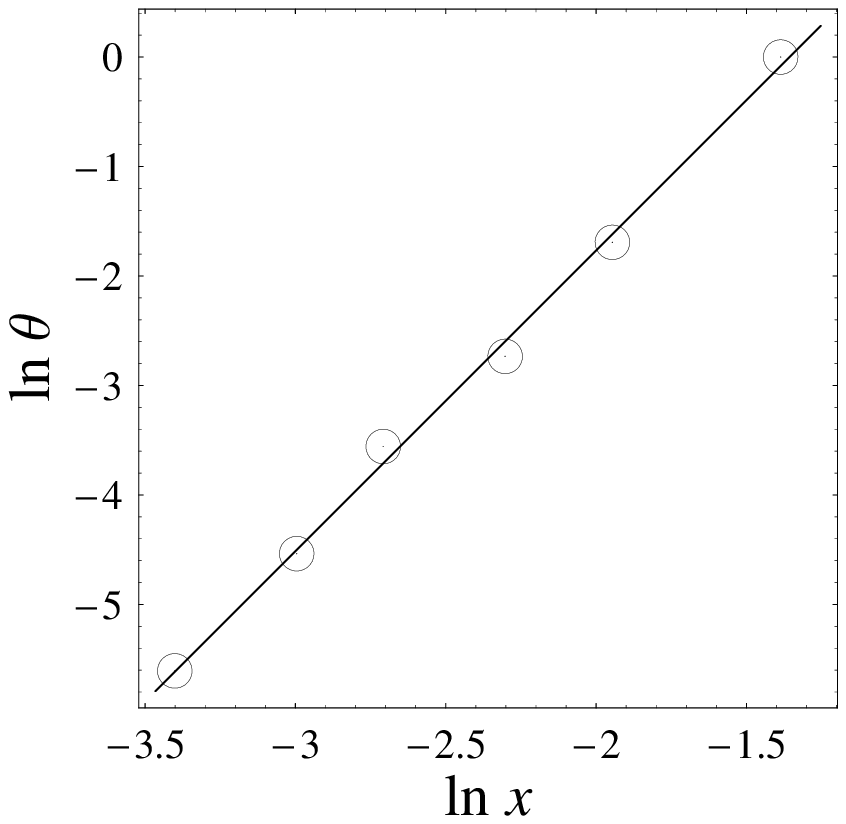}
\includegraphics*[width=1.65in]{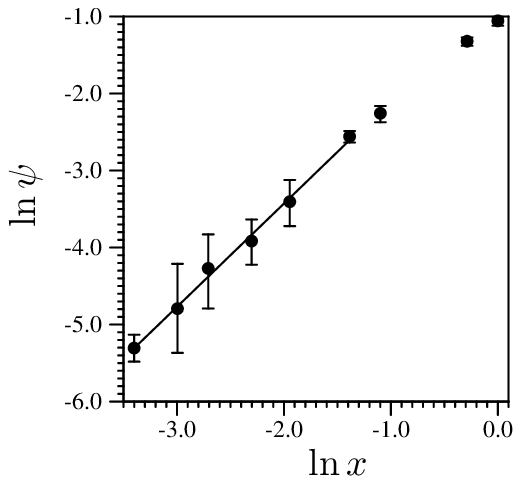}
\end{center}
\caption{Dependence of the rescaling factor $\theta(x)$ (left panel) and the
relevant scaling field $\psi$ (right panel) on $x$. The solid line in both
panels is the least-square linear fit.  Error-bars in the right panel are
$95\%$ confidence intervals.}
\label{fig:mu}
\end{figure}

$\alpha$ is calculated from the dependence of $\Lambda_c$ on $x$, where
$\Lambda_c = \bar{\Lambda}(W=W_c)$.  The right panel of Fig.~\ref{fig:phi0}
shows that $\alpha = 0.89(84,95)$.  Numerical verification of
Eq.~(\ref{eq:two-universal}) is done by a rescaling of the argument of
$\bar{\Lambda}/\Lambda_c$ for each $x$ individually by a factor $\theta(x)$
chosen such that all $\bar{\Lambda}/\Lambda_c$ collapse, and Fig.~\ref{fig:F}
shows the result.  Figure~\ref{fig:mu} shows that $\theta(x)\propto x^{\mu_1},
\mu_1=2.75(2.28,3.16)$ and $\psi(x)\propto x^{\mu_2}, \mu_2=1.34(22,46)$, which
gives $\mu=\mu_1+\mu_2=4.1\pm 0.6$.

The largest similarity with the one-parameter scaling Eq.~(\ref{eq:one}) is
achieved for $\alpha=1,\,\mu=1/\nu$, when Eq.~(\ref{eq:two-universal}),
expressed in terms of $\xi$, becomes
\begin{equation}\label{eq:xi-simple}
  \xi(L,w,x) \simeq(x L)\mathcal{F}_\pm((x L)^{1\over\nu}\psi(w)).
\end{equation}
$x$ influences the divergence of the correlation length of the infinite
system $\xi_\infty(w,x)\sim (|w|/x)^{-{1\over\nu}}$ for $L\gtrsim L_c(x)$.
The numerical results are not incompatible with $\alpha=1$, but nonetheless
strongly suggest $\mu\ne\alpha/\nu$.

The two-parameter scaling was obtained for samples of small $\delta L$,
including the single-layered case (Table~\ref{table}).  This allows one to
follow outflow RG trajectories from $L\to\infty$ back to the single-layered
finite $L$ case, and therefore provides an analytical mapping of the phase
diagram of $3d$ systems onto $2d$ systems, suggesting a possibility of
signatures of $3d$ AT also in $2d$ strongly-localized regime.  What exactly
those signature are, including a possibility of their experimental observation
in the mesoscopic regime, will be discussed elsewhere~\cite{Cerovski07c}.

Although the study carried here shows the possibility of AT in arbitrarily thin
$3d$ films (which should be of size $L\gtrsim L_c(x)$ in order to reduce the
large corrections to scaling due to the dimensional crossover) in the
strong-disorder regime, the critical conductance is strongly suppressed due to
the $x^\alpha$ prefactor.

The RG approach proposed seems to be a rather accurate description of AT and
allows the nontrivial extension of the one-parameter STL to two parameters. It
remains to be seen whether such an approach can be applicable to other problems
in physics, for instance in addition to the dimensional regularization.


\begin{thebibliography}{42}

\bibitem{Anderson58}
  P.~W. Anderson, {Phys. Rev.} \textbf{ {109}}, {1492} ({1958}).

\bibitem{Abrahams79}
 E.~Abrahams,  P.~W. Anderson,  D.~C. Licciardello and T. V. Ramakrishnan,
 Phys. Rev. Lett. \textbf{ {42}},  {673} ({1979}).

\bibitem{Vollhardt80}
 D.~Vollhardt and P. W\"olfle,
 Phys. Rev. Lett. \textbf{45}, 842 (1980);
 Phys. Rev. B \textbf{22}, 4666 (1980).

\bibitem{Langer66}
 J. S. Langer and T. Neal, Phys. Rev. Lett. \textbf{16}, 984 (1966).

\bibitem{Lee85}
 P.~A.~Lee and A.~D. Stone, Phys. Rev. Lett. \textbf{ {55}},  {1622} ({1985}).

\bibitem{Mesoscopic91}
  B.~L. Altshuler,  P.~A. Lee, and  R.~A. Webb, eds.,
  \emph{ {Mesoscopic Phenomena in Solids}} ({North-Holland},  {Amsterdam},
   {1991}).

\bibitem{Shapiro86}
 { {B.}~{Shapiro}}, {Phys. Rev. B} \textbf{ {34}},
   {4394} ({1986}); {Philos. Mag. B} \textbf{ {56}},
   {1031} ({1987});
 B.~L. Altshuler, V.~E. Kravtsov, and I.~V. Lerner,  {Phys. Lett. A}
  \textbf{ {134}},  {488} ({1989});
  B.~Shapiro, {Phys. Rev. Lett.} \textbf{ {65}},
   {1510} ({1990});
 { {K.}~{Slevin}}, { {P.}~{Marko\v{s}}}, {and}  { {T.}~{Ohtsuki}},
   {Phys. Rev. Lett.} \textbf{ {86}},
   {3594} ({2001});
   {Phys. Rev. B} \textbf{ {67}},
   {155106} ({2003});
   K.~Slevin,  Y.~Asada, and  L.~I. Deych,
   {Phys. Rev. B} \textbf{ {70}}, {054201} ({2004}).

\bibitem{Pichard81}
  J.~L. Pichard and  G.~Sarma,  
  {J. Phys. C} \textbf{34},  L127 (1981).

\bibitem{MacKinnon81}
  A.~MacKinnon and  B.~Kramer,
   {Phys. Rev. Lett.} \textbf{47}, 1546 (1981);
   {Z. Phys. B: Condens. Matter} \textbf{53}, 1 (1983).

\bibitem{MacKinnon94}
 { {A.}~{MacKinnon}},
   {J. Phys.: Condens. Matter} \textbf{ {6}},
   {2511} ({1994}).

\bibitem{Slevin99}
  K.~Slevin and  T.~Ohtsuki,
   {Phys. Rev. Lett.} \textbf{ {82}},
   {382} ({1999}).

\bibitem{Efetov83}
 { {K.~B.} {Efetov}},
   {Adv. Phys.} \textbf{ {32}},
   {53} ({1983}).

\bibitem{Hikami81}
 { {S.}~{Hikami}},
   {Phys. Rev. B} \textbf{ {24}},
   {2671} ({1981});
 { {V.~E.} {Kravtsov}}, {{I.~V.} {Lerner}}, and {{V.~I.} {Yudson}}, 
   Sov. Phys. JETP \textbf{ {67}},  {1441} ({1988}).
 { {F.}~{Wegner}},
   {Nucl. Phys. B} \textbf{ {316}}, {663} ({1989});
   {Z.  Phys. B: Condens. Matter} \textbf{ {78}},  {33} ({1990});
 { {E.}~{Bre\'zin}} {and} { {S.}~{Hikami}},
   {Phys. Rev. B} \textbf{ {55}}, {R10169} ({1992});
 { {S.}~{Hikami}},
   {Prog. Theor. Phys. Suppl.} \textbf{ {107}}, {213} ({1992}).

\bibitem{Potempa98}
  H.~Potempa and  L.~Schweitzer,
  J. Phys. Condens. Matter \textbf{10}, L431 (1998).

\bibitem{Cardy96}
  J.~Cardy,
  \emph{Scaling and Renormalization in Statistical Physics}
  (Cambridge University Press, Cambridge,  1996).

\bibitem{Cerovski07b}
 V.~Z.~Cerovski, to appear in Phys.~Rev.~B, 
 preprint No. cond-mat/0701306 (2007).

\bibitem{Cerovski07c}
 V.~Z.~Cerovski, (unpublished) (2007).

\end{thebibliography}
\end{document}